\documentclass[11pt]{article}
\title{Extending the first-order post-Newtonian scheme in
 multiple systems to the second-order
contributions to light propagation
\thanks{Supported by the National Natural Science Foundation
of China under Grant No.19835040 and 10273008}}
\author{XU Chong-ming, WU Xue-jun\\
 {\small Department of Physics, Nanjing Normal University,
 Nanjing 210097}}
\date{}
\begin{document}
%\songti
%\zihao{6} \ziju{0.07}
%\baselineskip=24pt
\maketitle

\begin{abstract}
%\begin{center}
%\begin{minipage}{5.2in}
%{\small
In this paper, we extend the first-order post-Newtonian scheme
in multiple systems presented by Damour-Soffel-Xu to the second-order
contribution to light propagation without changing the virtue
of the scheme on the linear partial differential equations
of the potential and vector potential.
The spatial components of the metric are extended to
second order level both in a global coordinates ($q _{ij}/ c^4$)
and a local coordinates ($Q _{ab}/ c^4$).
The equations of $q_{ij}$ (or $Q_{ab}$) are obtained from
the field equations.
The relationship between $q_{ij}$ and $Q_{ab}$ are presented in
this paper also.
%$q_{ij}$ (or $Q_{ab}$) do not influence on the linear equation of
%potential and vector equations.
In special case of the solar system (isotropic condition is applied
($q_{ij} = \delta _{ij} q $)), we obtain the solution of $q$.
Finally, a further extension of the second-order contributions in
the parametrized post-Newtonian formalism is discussed.

%\bigskip

%\end{minipage}
%\end{center}
\end{abstract}

\vspace*{1cm}

%\bigskip

Recently a series of space astrometric missions are proposed,
e.g. LISA(Laser Interferometer space Antenna),$^{[1]}$ GAIA (new
generation of HIPPARCOS(HIgh Precision PARallax Collecting
Satellite) for attaining angular accuracy of about $10^{-6}$ arcsec
in few hours of integration),$^{[2]}$ SIM(Space Interferometry
Mission),$^{[3]}$ DIVA(Double Interferometer for Visual
Astrometry),$^{[4]}$
FAME(Full-sky Astrometric Mapping Explorer),$^{[5]}$
ASTROD(Astrodynamical Space Test of Relativity
using Optical Devices)$^{[6]}$, Mini-ASTROD$^{[7]}$ and so on.
LISA and GAIA are taken as cornerstones of ESA(European Space Agency).
SIM and FAME have been planned by NASA(National Aeronautica and Space
Administration) and DIVA is in the progress in Germany.
Especially Mini-ASTROD is proposed by our Chinese
colleagues and is under Phase A Study.$^{[7]}$
In all of above missions
the light propagation has to be calculated to a more precise level, e.g.
in Mini-ASTROD the precision is about $10^{-13}$, which is the second order
post Newtonian(2PN) level (2PN precision for the sun is
$10^{-12} \left( \frac{R_\odot}{r} \right) ^2$, where $R_\odot$ and $r$
are the radius of the sun and the closest distance from light trajectory
to the center of the sun). Therefore the second order contribution to
light propagation is required to be considered.
On other hand, when the light pass nearby the
sun or any planet, the potential produced by $2^N$ multipole moment
is$^{[8]}$
$ V_{\rm mono} (l / r ) ^N $,
where $V _{\rm mono}$ is the potential created by the mass monopole,
$l$ the length deviated to monopole and $r$ the distance of the action.
For the quadrupole of the sun $(N=2)$
$ ( l / r )^2 < \epsilon ^2 $,
where $\epsilon ^2$ is the small parameter of post-Newtonian expansion
(nearby the sun $\epsilon ^2 \sim 10 ^{-6}$), $l$ can be estimated from
the oblateness of the sun$^{[9]}$. Therefore if we consider 2PN problem, the
1PN quadrupole has to be taken into account. When we calculate
relativistic multipole moments and astrodynamics (equations of motion for
plannets), we have to use the scheme
presented by Damour-Soffel-Xu$^{[10-13]}$, in which
a complete 1PN general relativistic celestial mechanics for N arbitrarily
composed and shaped, rotating deformable bodies is described.
Their scheme is widely abbreviated as DSX schme$^{[14]}$. In their
first paper$^{[10]}$, they give out the clear relation between potential $w$
$\&$ vector potential $w_i$ in a global coordinates
(they are directly related with the global metric tensor)
and $W$ $\&$ $W _a$ in a local coordinates. It is
so called ``Theory of Reference Systems", which can solve the difficulty
caused by what the multipole moments is calculated in every local
coordinates, but the influence of multipole moments on the light-ray
trajectory is considered in the global coordinates. Their paper
II$^{[11]}$ first obtained the complete and explicit laws of the motion
for N arbitrarily composed and shaped bodies in the form of an infinite
series which make the relativistic correction of multipole moments on
astrodynamics (Ephemeris) to be possible. Therefore DSX scheme is the best
framework for taking into account the 1PN multipole moments.
However DSX scheme is the first post-Newtonian (1PN) approximation
for particle motion.
The metric tensor in a global coordinates is written in the form

\begin{eqnarray}
g_{00} &=& - \exp  \left( - \frac{2w}{c^2} \right) + O(6) \, ,
 \label{1}\\
g_{0i} &=& -  \frac{4w_i}{c^3} + O(5) \, , \label{2}\\
g_{ij} &=& \delta _{ij} \exp \left( \frac{2w}{c^2} \right)
 + O(4) \, ,  \label{3}
\end{eqnarray}
and they satisfy
\begin{equation}\label{4}
 g_{00} g_{ij} = - \delta _{ij} + O(4) \, ,
\end{equation}
where $O(n)$ means $O(c^{-n})$. Eq.(4) is the conformal isotropic
condition (see Eq.(4.5) of Ref.[10]).
Potential $w$ and vector potential $w_i$
satisfy {\it linear} Partial Differential Equations(PDE).
The metric in a local coordinates has a similar form as in the global
coordinates, one only needs to change the small letters into
Capital letters.
It has been pointed out already, that DSX scheme has to be
extended to 2PN, then it is possible to discuss 2PN
contribution$^{[15]}$ to light propagation,
which means the Eq.(3)
to be extended to $O(6)$. 2PN contribution for lightray has been
discussed for a long time (early 80's) by Epstein and
Shapiro (1980)$^{[16]}$, Richter and Matzner(1982)$^{[17]}$, and
by others (later). But all of them consider only
in one global coordinate system,
therefore they can
not calculate the relativistic contribution from multipole moments which
should be calculated in the local coordinates. As we know the relativistic
theory of reference system established only after 1991$^{[10]}$,
although it is only 1PN approximation.

In this paper we will first extend the metric $g_{ij}$
(and $G_{ab}$) in DSX scheme to $O(6)$, and deduce the equations satisfied
by $w$, $w_i$ and $q_{ij}$ (also for $W$, $W_a$ and $Q_{ab}$ in every
local coordinates) from Einstein field equations, where we introduce new
quantities $q_{ij}$ (and $Q_{ab}$ in the local coordinates)
for metric $g_{ij}$.
Then the relationship between potentials
in the global and the local coordinates is deduced also.
After that, the special case of the solar system
($q_{ij}= \delta _{ij} q )^{[17]}$ is discussed, and an integrable
solution of $q$ is presented.
Finally we give a brief discussion.

All of symbols in this paper and conventions are the same as
in DSX scheme$^{[10-13]}$.
To apply DSX scheme to discuss the second
order contribution to light propagation,
$g_{ij}$ (or $G_{ab}$) should be extended
to $O(6)$ as we mentioned before.

The metric tensor in a local coordinate system now take the form
\begin{eqnarray}
G_{00} &=& - \exp \left( - \frac{2W}{c^2} \right) + O(6) \, ,
  \label{5}\\
G_{0a} &=& - \frac{4W_a}{c^3} + O(5) \, , \label{6}  \\
G_{ab} &=& \delta _{ab} \exp \left( \frac{2W}{c^2} \right)
         + \frac{Q_{ab}}{c^4}+ O(6) \, . \label{7}
\end{eqnarray}
The contravariant metric reads
\begin{eqnarray}
G^{00} &=& - \exp \left(  \frac{2W}{c^2} \right) + O(6) \, ,
   \label{8}\\
G^{0a} &=& - \frac{4W_a}{c^3} + O(5) \, , \label{9}  \\
G^{ab} &=& \delta _{ab} \exp \left( - \frac{2W}{c^2} \right)
         - \frac{Q_{ab}}{c^4}+ O(6) \, .  \label{10}
\end{eqnarray}
The metric tensor in the global coordinate system has a similar form,
just change $G_{\alpha\beta}$, $W$, $W_a$ and $Q_{ab}$ by
$g_{\mu\nu}$, $w$, $w_i$ and $q_{ij}$.

The relations between $w \, , \; w_i \, , \; q_{ij} $ and
$W \, , \; W_a \, , \; Q_{ab} $
will be studied in the following. In DSX scheme, since the metric in the
global coordinate system as well as in every local coordinate systems has
the similar form, so that, if we deduce any equation in one local
coordinate system, then we have it in every local coordinates
as well as in the global coordinates. Therefore we need only deduce the
field equation in a local coordinate system.
The spatial conformal isotropic condition Eq.(\ref{4}) is revised as

\begin{equation}\label{11}
G_{ab} G_{00} = - \delta _{ab} - \frac{Q_{ab}}{c^4} + O(6) \, .
\end{equation}
If we attribute $Q_{ab}/c^4$ to $O(4)$, it is just the form in DSX scheme.
Therefore  Eq.(\ref{11}) is an extension of Eq.(\ref{4}).
$Q_{ab}$ also can be taken as a spatial anisotropic
contribution in second order.

The metric tensor $g_{0i}$ is taken as $\epsilon ^3 $ here,
in fact it is only
as small as $\epsilon ^4$ because of the relatively slow rotation
rate in the sun $^{[17,18]}$.
To keep the uniformity with DSX scheme
therefore we take such formula of $g_{0i}$ as well as for $Q_{0a}$.

The Christoffel symbols can be calculated from the metric tensor
\begin{eqnarray*}
\Gamma ^0_{00} &=& - \frac{W_{,t}}{c^3} + O(5) \, , \\
\Gamma ^0_{0a} &=& - \frac{W_{,a}}{c^2} + O(6) \, , \\
\Gamma ^a_{00} &=& - \frac{W_{,a}}{c^2}
  + \frac{4 WW_{,a}}{c^4} - \frac{4W_{a,t}}{c^4}+ O(6) \, , \\
\Gamma ^0_{ab} &=& \delta _{ab} \frac{W_{,t}}{c^3}
  + \frac{4}{c^3} W_{(a,b)} + O(5) \, , \\
\Gamma ^a_{0b} &=& - \frac{4}{c^3} W_{[a,b]}
  + \frac{W_{,t}}{c^3} \delta _{ab} + O(5) \, , \\
\Gamma ^a_{bc} &=&  \frac{1}{c^2} (\delta _{ab} W_{,c}
  + \delta _{ac} W_{,b} - \delta _{bc}W_{,a} )
  +  \frac{1}{2c^4} ( Q_{ab,c}
  + Q_{ac,b} - Q_{bc,a} ) + O(6) \, ,
\end{eqnarray*}
where two indices enclosed in a parentheses is denoted as symmetrization,
and in a square bracket means antisymmetrization.

Then we can deduce the Ricci tensor
\begin{eqnarray}
R^{00} &=& - \frac{\nabla ^2 W}{c^2} - \frac{1}{c^4}
  \left( 3\partial _t \partial _t W
  + 4\partial _t \partial _d W_d \right) + O(6) \, ,  \label{12} \\
R ^{0a} &=& - \frac{2}{c^3} \left( \nabla ^2 W _a
  - \partial _a \partial _d W_d
  - \partial _t \partial _a W \right) + O(5) \, , \label{13}\\
R ^{ab} &=& - \frac{1}{c^2} \delta _{ab} \nabla ^2 W
  + \frac{1}{c^4} \left[ 4 \delta _{ab} W \nabla ^2 W
  +  \delta _{ab} W_{,tt}
  + 4 W_{(a,b),t} - 2W_{,a}W_{,b} \right. \nonumber \\
  && \left. + \frac{1}{2} ( Q_{ad,bd} + Q_{bd,ad}
  - Q_{ab,dd} - Q_{dd,ab}) \right] + O(6) \, , \label{14}
\end{eqnarray}
and the scalar curvature
\begin{equation}
R = - \frac{2}{c^2} \nabla ^2 W + \frac{1}{c^4}
  ( 4W \nabla ^2 W + 6 W_{,tt} + 8 W_{a,at}
  - 2 W_{,a} W_{,a} + Q_{ab,ab} - Q_{aa,bb} ) +O(6) \, . \label{15}
\end{equation}

From Einstein field equation we can derive equations of $W$, $W_a$
and $Q_{ab}$
\begin{eqnarray}
&& \nabla ^2 W + {1 \over c^2} ( 3W_{,tt}
  + 4 \partial _t \partial _a W_a) = - 4 \pi G \Sigma
  + O(4) \, , \label{16}\\
&& \nabla ^2 W_a - \partial _a \partial _b W_b
  -  \partial _t \partial _a W = - 4 \pi G \Sigma ^a
  + O(2) \, , \label{17} \\
&& -2 \delta _{ab} W _{,tt} + 4( W_{(a,b),t}
  - \delta _{ab} W _{(d,d),t} ) -2 W_{,a} W_{,b}
  + \delta _{ab} W_{,d}W_{,d} \nonumber \\
&& \quad + \frac{1}{2} ( Q_{ad,bd} + Q_{bd,ad}
  - Q_{ab,dd} - Q_{dd,ab} - \delta _{ab} Q_{dc,dc}
  + \delta _{ab} Q_{dd,cc} ) = 8 \pi G T ^{ab} \, , \label{18}
\end{eqnarray}
where
$ \Sigma = ( T^{00} + T ^{aa} ) / c^2 $ and
$ \Sigma ^a = T^{0a} /c $ (the same definition in DSX scheme).
Eq.(\ref{16}) and (\ref{17}) are linear PDE, which can be solved in
certain suitable gauge conditions, then Eq.(\ref{18}) will be solved also.
We shall point out that,
$T^{ab}$ can be expressed by $\Sigma$ and $\Sigma ^a$
(to see Eq.(A13) of Ref.[19]).
Therefore it is self-consistent within the framework of
DSX scheme.
Similar field equations in the global coordinate system can be
obtained easily by substituting capital letters by small
letters.

From Eqs.(16)-(17) we see that $Q_{ab}$ does not appear
in the field equations of $W$ and $W_a$, and they are the same as ones
in DSX scheme (keeping linear equations). The solutions of $W$ and
$W_a$ related to relativistic multipole moments are still valid as
before in DSX scheme.

Now, we shall discuss the transformation relations of the potential and
the vector potential from the global
coordinate system to a local coordinate system and vice versa.
The coordinate transformation law reads
\begin{equation}\label{19}
g^{\mu\nu} = \frac{\partial x^\mu}{\partial X^\alpha}
  \frac{\partial x^\nu}{\partial X^\beta} G ^{\alpha\beta} \, .
\end{equation}
Considering $g^{00}$, we obtain
\begin{equation}\label{20}
w=\left( 1 + \frac{2V^2}{c^2} \right) W
  + \frac{4}{c^2} V^a W_a
  + \frac{c^2}{2} \ln ( A^0_0 A^0_0
  - A^0_a A^0_a ) + O(4) \, ,
\end{equation}
where $A^\mu_{\alpha} = \partial x ^\mu / \partial X ^\alpha$,
$ V ^a= R ^a _i  v^i$ ($R^a _i$ is a slowly changing rotation matrix
and $v^i $ is the coordinate three-velocity of the central world line
measured in the global coordinate system). \\
For $g^{0i}$, it turns out to be
\begin{equation}\label{21}
w _i = v^i W + R^i_a W_a
  + \frac{c^3}{4} ( A^0_0 A^i_0
  - A^0_a A^i_a ) \, .
\end{equation}

In Eqs.(\ref{20}) and (\ref{21}), $Q_{ij}$ does not enter these equations,
and the relations between $w$, $w_i$ and $W$, $W_a$
are the same as in DSX scheme (to keep linear relation).
As we know, all of the discussion about the theory of reference systems
are based on the linear PDE of potentials (Eq.(\ref{16}), (\ref{17}))
and linear relation between potentials in the global coordinates and
in the local coordinates ((\ref{20}) and (\ref{21})).
Since these equations are the same as before (in DSX scheme),
the theory of reference system in DSX scheme (for potential and
vector potential) is valid in this paper also. \\
Finally, for $g^{ij}$, we get
\begin{eqnarray}
q _{ij} &=& -2W ( 2 V^2 \delta _{ij}  - R^i_a  R^j_b
  V^a V^b ) - 8 V^a W _a \delta _{ij}
  + 8 R_a ^{(i} R ^{j)} _b W ^a V ^b
  + R^i_a R^j_b Q_{ab} \nonumber \\
  && + 2 W c^2 (A^i _a A ^j _ a - \delta _{ij})
  + c^4 \{  A^i_0 A^j_0 - A^i_a A^j_a
  + \delta _{ij} [ 1 - \ln ( A^0_0 A^0_0
  - A^0_a A^0_a )  ] \} \, . \label{22}
\end{eqnarray}
Here we should emphasize that Eq.(\ref{22}) has only formally given for
completeness. When we consider the problem of the light propagation
in the solar system Eq.(\ref{22}) is unnecessary to be considered which
we will explain later.

To discuss the gravitational influence on light propagation for
Mini-ASTROD$^{[7]}$ $(10 ^{-13})$ we consider the solution in the solar
system now. Potential $W$ and vector potential $W_a$ can be obtained in
the same as in Ref.[10] by means of B-D multipole moments$^{[20]}$
in every local coordinates.
Therefore we shall not repeat it again. $Q_{ab}$ is
rather complex. Fortunately for the problem of the light propagation
in the solar system it becomes much simple. Since the velocity of the
relative motion is small inside the sun ($v^2 / c^2 < \epsilon ^2$),
especially the shape of the sun is close to a monopole, therefore if
the origin of the coordinate system is taken as the center of the solar
mass (dipole always equal to zero), the quadrupole terms (and higher
multipole moments) in 2PN contribution on $Q_{ab}$ can be ignored, i.e.
$Q_{ab} = \delta _{ab} Q$, as well as in the global coordinates of the sun
\begin{equation}\label{23}
q_{ij} = \delta _{ij} q \, .
\end{equation}
As for other local coordinates of every planets 2PN contribution is too
small ($10^{-19}-10^{-18}$) to be calculated.

Substituting Eq.(\ref{23}) into Eq.(\ref{18}) and taking PN gauge
($ 3\partial _t W + 4 \partial _a W_a = O(2)$), $Q$ satisfies the
equation as following:
\begin{equation}\label{24}
\left( Q + {1 \over 2} W^2 \right) _{,dd} = 8\pi G \left( T^{aa}
  - \frac{W \Sigma}{2} \right) \, .
\end{equation}
The right hand side of above equation has a compact support
and the solution of Eq.(\ref{24}) is
\begin{equation}\label{25}
Q = -  \int \frac{2G ( T^{aa}- W \Sigma /2)}{r} d^3 X
  - {1 \over 2} W ^2 \, .
\end{equation}
From  Eq.(\ref{25}) it is clear that $Q$ itself does not have compact
support neither is linear in $W$ (because of the last term).

In the global coordinates of the solar system, we have a similar equation
and solution. Since there is very small difference between the global
coordinates of the solar system and the local coordinates of the sun
(additional $10^{-3}$ in $c^{-4}$ level) we can always ignore the
distinction between
$q$ and $Q$ of the sun. This is the reason why Eq.(\ref{22}) is
unnecessary to be considered.

We have extended DSX scheme to 2PN level approximation.
In the solar system we obtain the
solution of $q$ and $Q$. Our extension keep the main virtue of DSX scheme
(linear equations of the potential and the vector potential and linear
relationship between $w$, $w^i$ and $W$, $W_a$).
Here we discuss further on our results.

This extension may also be used in parametrized post-Newtonian
PPN formalism, which would be found a widely use for the relativistic
experiments and observation$^{[21]}$. For DSX scheme (1PN) the PPN
extension is written
as$^{[22]}$
\begin{eqnarray}
g_{00} &=& -1 + \frac{2w}{c^2} - \frac{2}{c^4} \beta w^2
  + O(6) \, , \\
g_{0i} &=&  - \frac{2(1 + \gamma )}{c^3} w^i
  + O(5) \, , \\
g_{ij} &=& \delta _{ij} \left( 1 + \frac{2\gamma }{c^2} w \right)
  + O(4) \, .
\end{eqnarray}
Our extension for parametrized second post Newtonian (PP$^2$N)
formalism for the solar system is taking the form
\begin{eqnarray}
g_{00} &=& -1 + \frac{2w}{c^2} - \frac{2}{c^4} \beta w^2
  + O(6) \, , \\
g_{0i} &=&  - \frac{2(1 + \gamma )}{c^3} w^i
  + O(5) \, , \\
g_{ij} &=& \delta _{ij} \left( 1 + \frac{2\gamma }{c^2} w
  + ( 2 \gamma ^2 - \frac{\delta}{2} ) \frac{w^2}{c^4}
  - \frac{2 \epsilon }{c^4} \int \frac{G \overline{T}^{ss}}{r} d^3 x
  + \frac{\xi}{c^4} \int \frac{w\sigma}{r} dx^3 \right)
  + O(6) \, ,
\end{eqnarray}
where $\overline{T} ^{ij}$ is the energy-momentum tensor in the global
coordinates.
When $\gamma = \beta = \delta = \varepsilon = \xi = 1$, then the
metric tensor return to the case in general relativity. If we ignore all
of $1 / c^4$ terms in $g_{ij}$, our results agree with the one of
Klioner and Soffel.$^{[22]}$

%{\bf Aknowledgement}
We would like to thank Prof. Ni Wei-tou and Prof. Tao Jinhe for the useful
discussion.

%%%%%%%%%%%%%%%%%%%%%%%%%%%%%%%%%%%%%%%%%%%%%%%%%%%%%%%%%%%%%%%%%%%%%

\bigskip\noindent
{\large \bf REFERENCES}

\noindent
{\small
\noindent
{}[1] LISA study team  2000 LISA: A Cornerstone Mission for the
Observation of Gravitational Wave \\
 {}\hspace*{0.5cm} ESA System and Technology Study Report ESA-SCI
  11 July 2000 \\
{}[2] Perryman M A C et al. 2001 Astron. \& Astrophys. {\bf 369}
  339\\
{}[3] Boden A, Unwin S and Shao M 1997 Proc. of ESA Symposium 402
``Hipparcos Venice 97", \\
 {}\hspace*{0.5cm} ed. Perryman M A C and Bernacca P L (ESA Pub. Division
 Noordwijk The Netherlands) 789 \\
{}[4] R\"oser S, Bastian U and  de Boer K S 1997 Proc. of ESA Symposium 402
 ``Hipparcos Venice 97" \\
 {}\hspace*{0.5cm} ed. Perryman M A C and Bernacca P L
 (ESA Pub. Division Noordwijk The Netherlands) 777 \\
{}[5] Triebes K J et al. 2000
  SPIE {\bf 4013} 482 \\
{}[6] Ni W-T et al. 2002 Intern. J. Modern Phys. D {\bf 11} No.7
 947 \\
{}[7] Ni W-T et al. 2002 Intern. J. Modern Phys. D {\bf 11} No.7,
 1035 \\
{}[8] Xu C, Wu X and Sch\"afer G 1997 Phys. Rev. D {\bf 55} 528. \\
{}[9] Will C 1993 ``Theory and experiment in gravitational physics"
(Cambridge Univ. Press UK)\\
 {}\hspace*{0.5cm} pp. 181-182. \\
{}[10] Damour T, Soffel M and Xu C 1991 Phys. Rev. D {\bf 43} 3273\\
{}[11] Damour T, Soffel M and Xu C 1992 Phys. Rev. D {\bf 45} 1017\\
{}[12] Damour T, Soffel M and Xu C 1993 Phys. Rev. D {\bf 47} 3124\\
{}[13] Damour T, Soffel M and Xu C 1994 Phys. Rev. D {\bf 49} 618\\
{}[14] Resolutions B1 of the XXIVth General Assembly, IAU
        Informational Bulletin {\bf 88}\\
 {}\hspace*{0.5cm}(the Astronomical Society of Pacific, San Francisco, 2001).\\
{}[15] Yi Z H 2002 J Yunnan Observatory {\bf 3} 9 \\
{}[16] Epstein R and Shapiro I I 1980 Phys. Rev. D {\bf 22} 2947\\
{}[17] Richter G W and Matzner R A 1982 Phys. Rev. D {\bf 26}
   1219 \\
{}[18] Richter G W and Matzner R A 1981 Astrophy. and Space Sci.
   {\bf 79} 119 \\
{}[19] Xu C and Wu X 2001 Phys. Rev. D {\bf 63} 064001 \\
{}[20] Blanchet L and Damour T 1989 Ann. Inst. Henri Poincare {\bf 50}
  377\\
{}[21] Qin B, Wu X and Zou Z 1997 Chin. Phys. Lett. {\bf 14}, 155 \\
{}[22] Klioner S and Soffel M 2000 Phys. Rev. D {\bf 62} 024019

}

\end{document}